\documentclass[a4paper,aps,pre,floatfix,nofootinbib,twocolumn,showpacs,bibtex]{revtex4-1}

\usepackage[pdftex]{graphicx}
\usepackage{latexsym,amsmath,verbatim}
\usepackage{color}
\usepackage{rotating}
\usepackage{fullpage}
\usepackage{multirow}
\usepackage[english]{babel}

\newcommand{\be}{\begin{equation}}
\newcommand{\ee}{\end{equation}}

\begin{document}

\title{Rescaling citations of publications in Physics}

\author{Filippo Radicchi}
\affiliation{Chemical and Biological Engineering, Northwestern University,
2145 Sheridan Road, Evanston, IL 60208, US}

\author{Claudio Castellano}

\affiliation{Istituto dei Sistemi Complessi (CNR-ISC), UOS Sapienza,
P.le A. Moro 2, I-00185 Roma, Italy}

\affiliation{Dipartimento di Fisica, ``Sapienza''
Universit\`a di Roma, P.le A. Moro 2, I-00185 Roma, Italy}

\date{\today}

\begin{abstract}
We analyze the citation distributions of all papers published
in Physical Review journals between $1985$ and $2009$.
The average number of citations received by papers published in a given
year and in a given field is computed. Large variations are found,
showing that it is not fair to compare citation numbers across
fields and years.
However, when a rescaling procedure by the average is used, it is possible to
compare impartially articles across years and fields.
We make the rescaling factors available, for use by the readers.
We also show that rescaling citation numbers by the number of publication
authors has strong effects and should therefore be taken into account
when assessing the bibliometric performance of researchers.
\end{abstract}


\maketitle

\section{Introduction}
Despite its many shortcomings, citation analysis is increasingly
used as a quantitative tool to evaluate research performance,
ranging from the single publication up to the level of individual
researchers, groups, departments, institutions and
countries~\cite{price65,egghe90,moed05,
hirsch05,bar08,stringer08,petersen10, stringer10}.
The underlying idea is that the number of citations measures the
impact of a publication and it is thus a proxy for the quality and
importance of the scientific work described in it.
This assumption is highly questionable and the empirical evidence
revealing striking counterexamples to its general validity is 
substantial~\cite{macroberts96,adler08, bornmann08}.
In principle, a careful, unbiased review by peers with no conflict
of interest would be a better procedure to evaluate research quality
and impact (although many distortions can occur also in this
case~\cite{bornmann11}).
Nevertheless, for several reasons that go much beyond the scope of this work
and will not be discussed here,
the use of quantitative bibliometric measures is not going to diminish
in the future.
It is much more realistic and effective trying to correct the inadequacies
of how citation analysis is performed, rather
than self-deceptively hope that citation analysis will lose importance
in the future~\cite{lane10}.

In this paper, we tackle two of the more obvious problems in the
use of the number of citations to measure the impact of scientific 
publications: 1) papers in some fields are typically
more cited than papers in other fields; 2) old papers naturally
tend to have more citations than more recent ones. 
These facts are part of the common wisdom of people working in science,
yet it is common to see comparisons between the number of cites of papers
dealing with completely different topics or published in different
decades.
However, as we will show, one should not take for granted that a paper
which has been cited twice as much as another in a different field
has actually had a larger impact.
The most natural way to get rid of (or at least alleviate) these problems
is the use of relative indicators, i.e. the normalization of the number of
citations by some average over a suitable reference set.
The use of relative indicators has been proposed long ago~\cite{schubert86}
and used in various contexts, but often in questionable
ways~\cite{leydesdorff10, waltman10}
and without an empirical check that it actually solves the problem.
Recently we have shown that such a normalization strongly reduces
biases when papers from entirely different disciplines, spanning all
fields of science, are compared~\cite{radicchi08, castellano09}.
Here we present a similar analysis applied to publications of the American
Physical Society (APS).
\begin{table*}
\begin{tabular}{|c l|}
\hline
PACS & Description \\
\hline
$00$ & General \\ 
$10$ & The Physics of Elementary Particles and Fields \\ 
$20$ & Nuclear Physics \\ 
$30$ & Atomic and Molecular Physics \\ 
$40$ & Electromagnetism, Optics, Acoustics, Heat Transfer, Classical Mechanics, 
and Fluid Dynamics \\ 
$50$ & Physics of Gases, Plasmas, and Electric Discharges \\ 
$60$ & Condensed Matter: Structural, Mechanical and Thermal Properties \\ 
$70$ & Condensed Matter: Electronic Structure, Electrical, Magnetic, and Optical
 Properties \\ 
$80$ & Interdisciplinary Physics and Related Areas of Science and Technology \\ 
$90$ & Geophysics, Astronomy, and Astrophysics \\ 
\hline
\end{tabular}
\caption{The ten categories considered in this study based on the
first digit of the first field of the PACS numbers}
\label{codes}
\end{table*}
With respect to Ref.~\cite{radicchi08, castellano09} we consider a
more fine-grained level of categorization, since the different
groups we deal with are all within the realm of Physics.
Fields are identified using the PACS (Physics and Astronomy
Classification Scheme~\cite{PACS}) number scheme.
We show first that
relevant differences in citation patterns show up among different fields.
The average number of citations in some field can be up to three
 times larger than
for papers in another field.
Analogously, we show that papers published in $1985$ are cited in some fields
even 15 times more than papers published in $2009$.
We then show that introducing a suitable relative indicator, the
distributions of such indicator become essentially
independent from the field or the year of publication, so that the
relative indicator can be used as an unbiased measure of impact.
We corroborate these results by showing that ranking papers
based on the raw number of citations leads to large biases in favor
of some PACS codes, while ranking using the relative indicator
allows a fair comparison among different fields.
We make available, for reference, a table with the average number
of citations for each category and each publication year since 1985.
We finally present further evidence of the relevance of the rescaling
procedure by considering all authors who published in APS journals
between $1985$ and $2006$. There is of course a correlation between
the number of raw citations of an author and its/her relative
indicator, but large deviations are possible. The problem
is even more remarkable when the relative indicator includes a
rescaling with the number of authors of each paper, so that the
number of citations of a paper are equally split among all co-authors.
This highlights that rescaling cites with the number of authors
is a crucial issue in citation analysis.

\section{Data sets}
\begin{figure}
  \begin{center}
    \includegraphics*[width=0.45\textwidth]{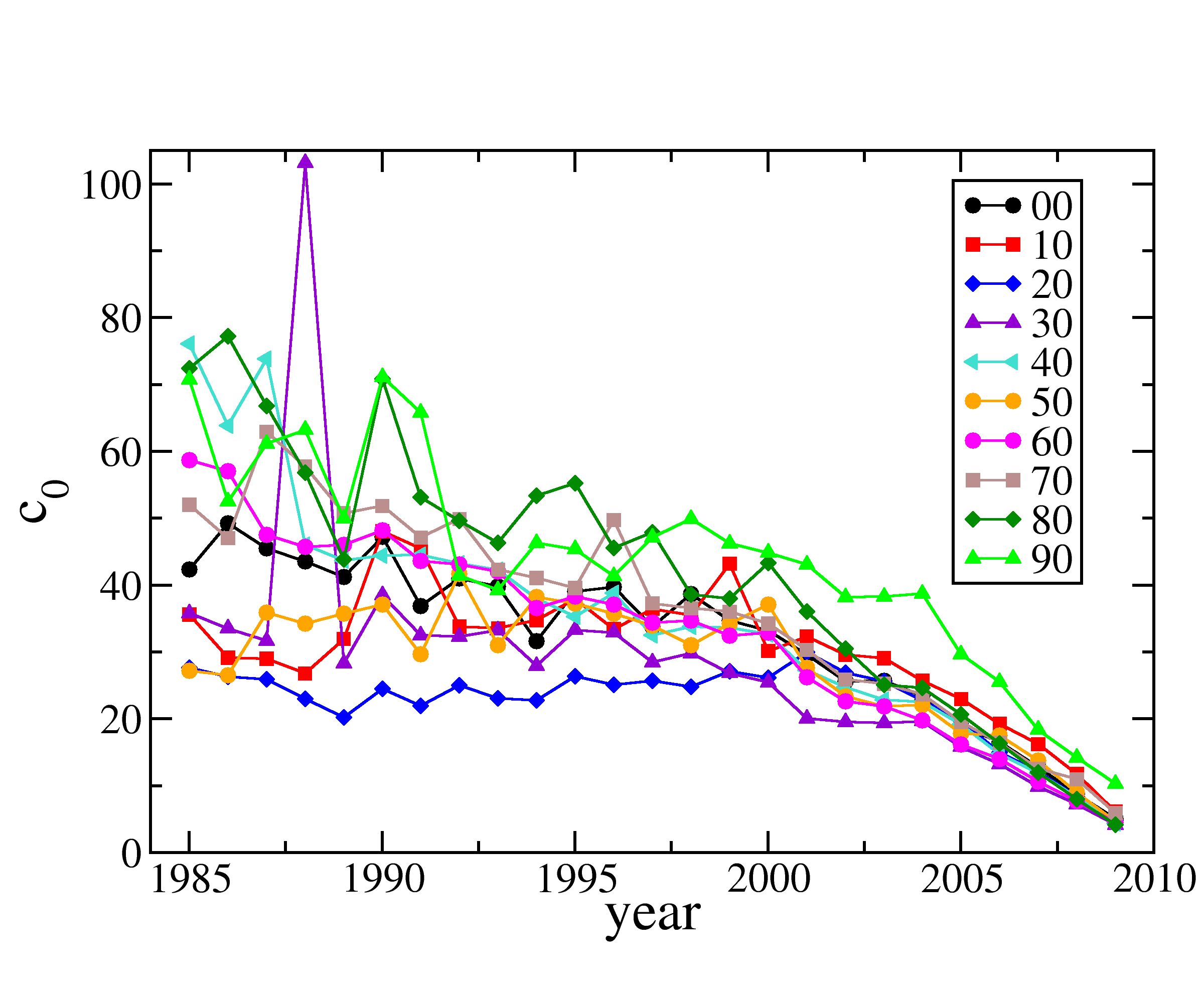}
  \end{center}
  \caption{(Color online) Value of the average number of cites per paper $c_0$ {\it vs}
publication year for the different categories.}
  \label{figc0}
\end{figure}
We consider all papers published in journals of the American Physical 
Society (APS, {\tt www.aps.org})
from $1985$ to $2009$. We restrict our analysis only
to standard research publications (Letters, Rapid communications, Brief
Reports and Regular Articles) 
and exclude other type of published material (Editorials, Reviews,
Comments, Replies and Errata)
which may show distinct citation patterns.
The journals considered in our analysis are:
Physical Review Letters,
Physical Review A, Physical Review B,
Physical Review C, Physical Review D  and Physical Review E.
APS journals represent the most important publication outlets
in Physics and cover all sub-fields of this discipline.
They therefore represent an optimal benchmark for the study 
of citation patterns of publications within Physics~\cite{redner05}.
The first year of the temporal range considered
has been selected because in $1985$ the PACS coding started to be
systematically used.
We consider only papers classified according to the PACS codes, which
are the vast majority ($>95 \%$ between $1985$ and $2009$) 
of all papers published in APS journals.
PACS numbers are attributed to papers by authors themselves.
This guarantees an optimal classification into fields, 
overcoming the nontrivial problem of attributing,
a posteriori, papers to fields~\cite{leydesdorff10, boyack05}.
PACS codes are composed of three fields $XX.YY.ZZ$,
where the first two are numerical (two digits each) and the third is
alphanumerical. For our purpose we consider only the first digit of
the $XX$ code, which provides a classification into
very broad categories (see Table~\ref{codes}).
Hence, for example, two papers with PACS codes
$05.70.Ln$ and $02.50.2r$  both belong to the category $00$, while a 
paper with PACS number $64.60.Ht$ is part of the category $60$.
In general, authors assign to a paper two
or three PACS numbers. In our analysis
we classify papers only according to their principal PACS number.
For each paper the number of cites is obtained from the
WebOfScience (WOS, {\tt www.isiknowledge.com}) database, 
hence including also citations from all
other non-APS publications included in the WOS database.
The data collection was performed on Dec. $15$, $2010$.
In our analysis we consider only papers which
have received at least one citation until the
above mentioned date.

\section{The relevance of the problem. Mean values and distributions}

In Fig.~\ref{figc0} we report, for each category,
the number $c_0$ of citations received on average by each paper
as a function of the publication year. 
It turns out that there are quite large differences in the values of
$c_0$ depending both on the category considered and on the year.
In particular, articles published in the same year in a given field
can be cited on average up to three times more than in another field
(e.g. PACS $40$ {\it vs} PACS $50$ in $1985$).
The average number of citations received tends to grow as
older publications are considered, with rather limited fluctuations, 
with the notable exception of the very high peak visible in $1988$
for PACS $30$, due to the extraordinary popularity (more than $30\,000$ 
citations so far) of a single article~\cite{verypopular}.
In general, it is possible to see that the differences 
between the $c_0$s of 
papers published under the same PACS number
but in different years  can be very large:
for PACS $80$ for example, $c_0\simeq 80$ for
articles published in $1986$ but only $c_0\simeq 8$
for publications of $2008$.

These data indicate that, although the variations are less pronounced with
respect to cross-discipline comparisons~\cite{radicchi08, castellano09},
there are important caveats
when trying to compare the number of citations accrued
by articles published in APS journals with different PACS codes or
publication years.

\begin{figure}
  \begin{center}
    \includegraphics*[width=0.45\textwidth]{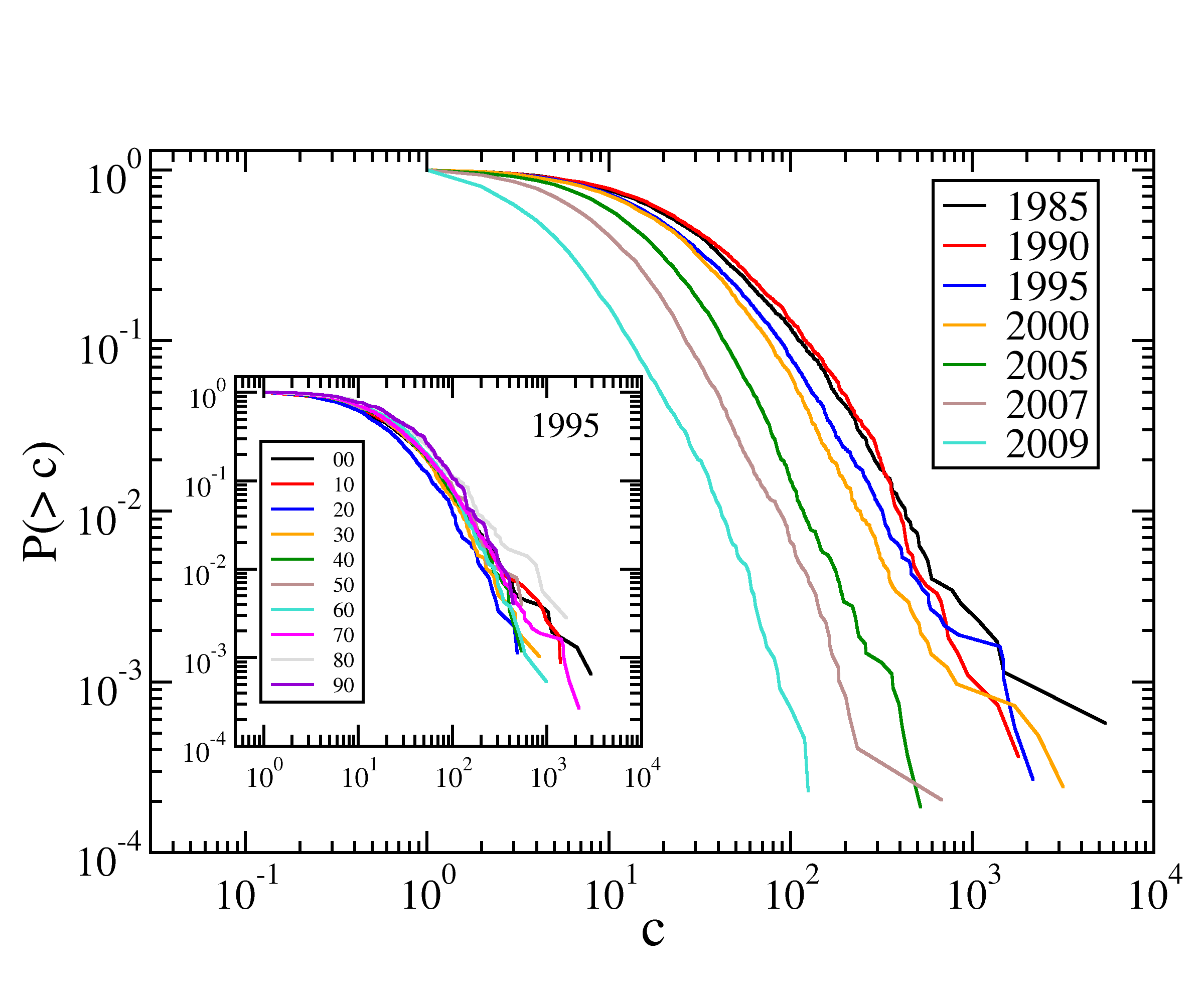}
  \end{center}
  \caption{(Color online) In the main panel, we plot the probability $P\left( >c \right)$
to observe a paper with more than $c$ citations   for category PACS $70$ and
years, from top to bottom: $1985$ (black), $1990$ (red), 
$1995$ (blue), $2000$ (orange), 
$2005$ (green), $2007$ (brown) and $2009$ (turquoise). 
In the inset, we plot $P\left( >c \right)$ for papers
published in the same year $1995$ but in different PACS
categories: $00$ (black), $10$ (red), 
$20$ (blue), $30$ (orange), 
$40$ (green), $50$ (brown), $60$ (turquoise), 
$70$ (magenta), $80$ (gray) and $90$ (violet).}
  \label{distr_unscaled}
\end{figure}

The same problem is found when one does compare not only the average values
but the full distributions.
For example, in Fig.~\ref{distr_unscaled} we plot the
cumulative distribution functions (cdf) 
of the number of citations for articles in the category $70$
for various publication years.
It is evident that over the years the distributions change considerably.
For instance, more than $200$ citations is not an uncommon result for a paper
published in $1985$, while it is an extraordinary achievement for one
published in $2005$.
Similarly, citation distributions are incompatible also when one considers
the same publication year but different PACS categories, as shown
in the inset of Fig.~\ref{distr_unscaled} for year $1995$.

\section{Relative citation rates and unbiased ranking}
\begin{figure}
  \begin{center}
    \includegraphics*[width=0.45\textwidth]{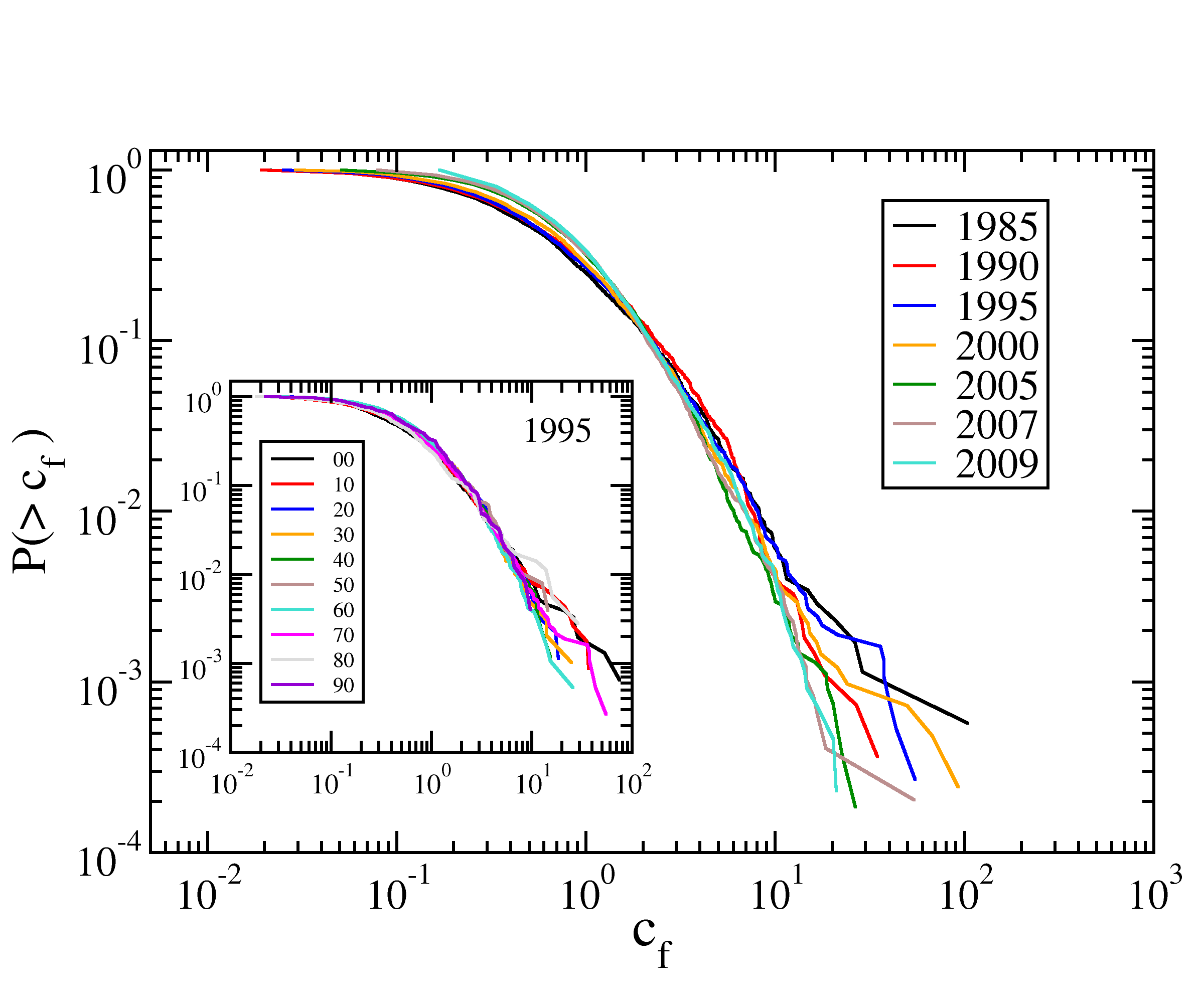}
  \end{center}
 \caption{(Color online) In the main panel, we plot the probability $P\left( >c_f \right)$
to observe a paper with relative citation count larger than $c_f$ for category PACS $70$ and
years $1985$ (black), $1990$ (red), $1995$ (blue), $2000$ (orange),
$2005$ (green), $2007$ (brown) and $2009$ (turquoise).
In the inset, we plot $P\left( >c_f \right)$ for papers
published in the same year $1995$ but in different PACS
categories: $00$ (black), $10$ (red), 
$20$ (blue), $30$ (orange), 
$40$ (green), $50$ (brown), $60$ (turquoise), 
$70$ (magenta), $80$ (gray) and $90$ (violet).}
  \label{distr_scaled}
\end{figure}

The most natural solution to the problem pointed out in the previous section
is the use of relative citation numbers. Let us define the ratio
\be
c_f = \frac{c}{c_0}.
\ee
\begin{figure*}
\begin{center}
\includegraphics[width=0.7\textwidth]{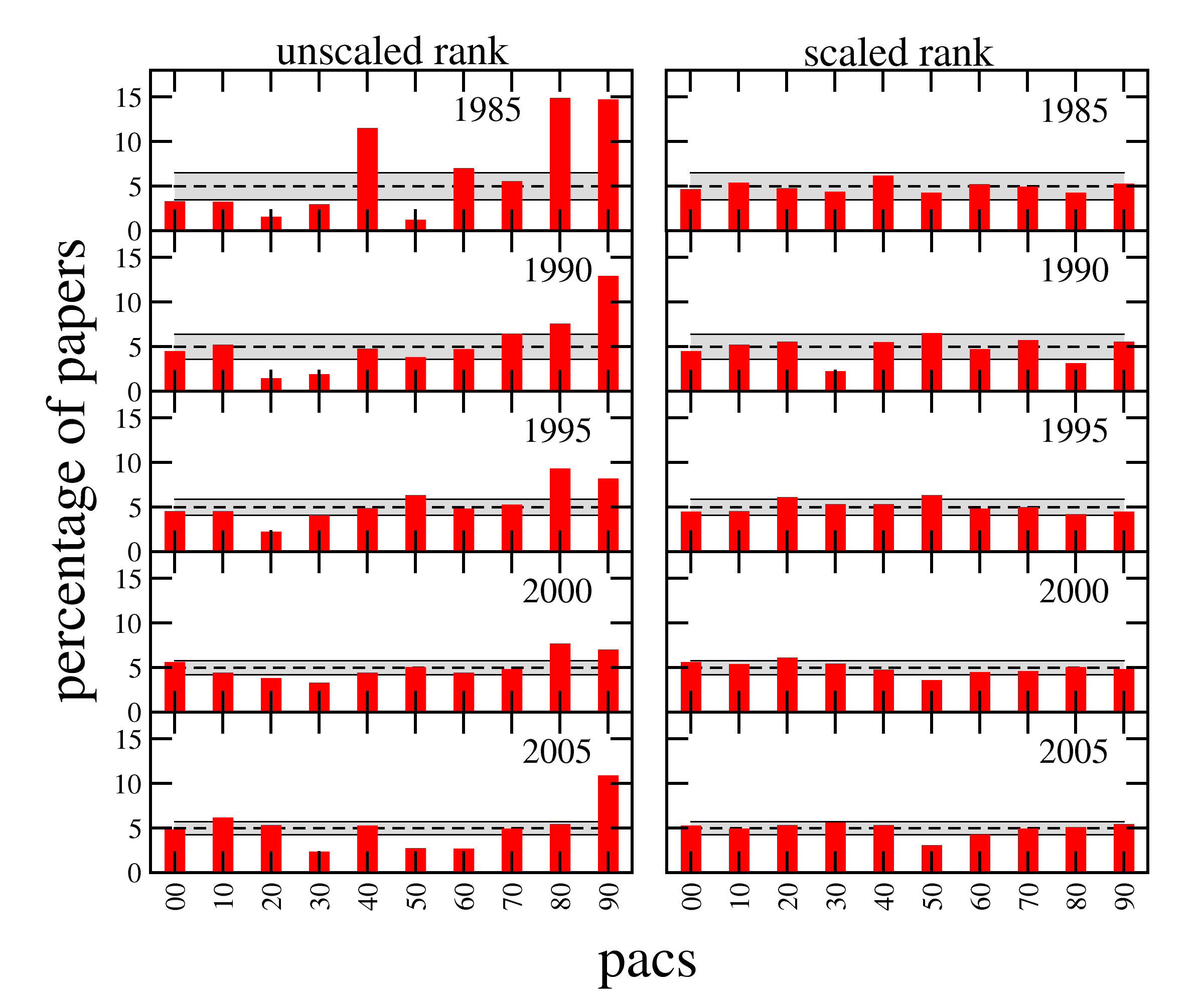}
\end{center}
\caption{(Color online) Histograms representing the percentage of papers belonging
to the top $5\%$ of the global rank performed according to $c$ (left
column) or $c_f$ (right column). In each
panel we consider only papers
published in a given year and plot the percentage
of papers with given PACS category belonging to
the top $5\%$ of the global rank. 
Black dashed lines represent the theoretically
expected values (in case of a fair ranking) of the mean, while
the gray areas cover the regions corresponding to the
mean $\pm$ one standard deviation.
}
\label{z=5}
\end{figure*}
\begin{figure}
\begin{center}
\includegraphics[width=0.45\textwidth]{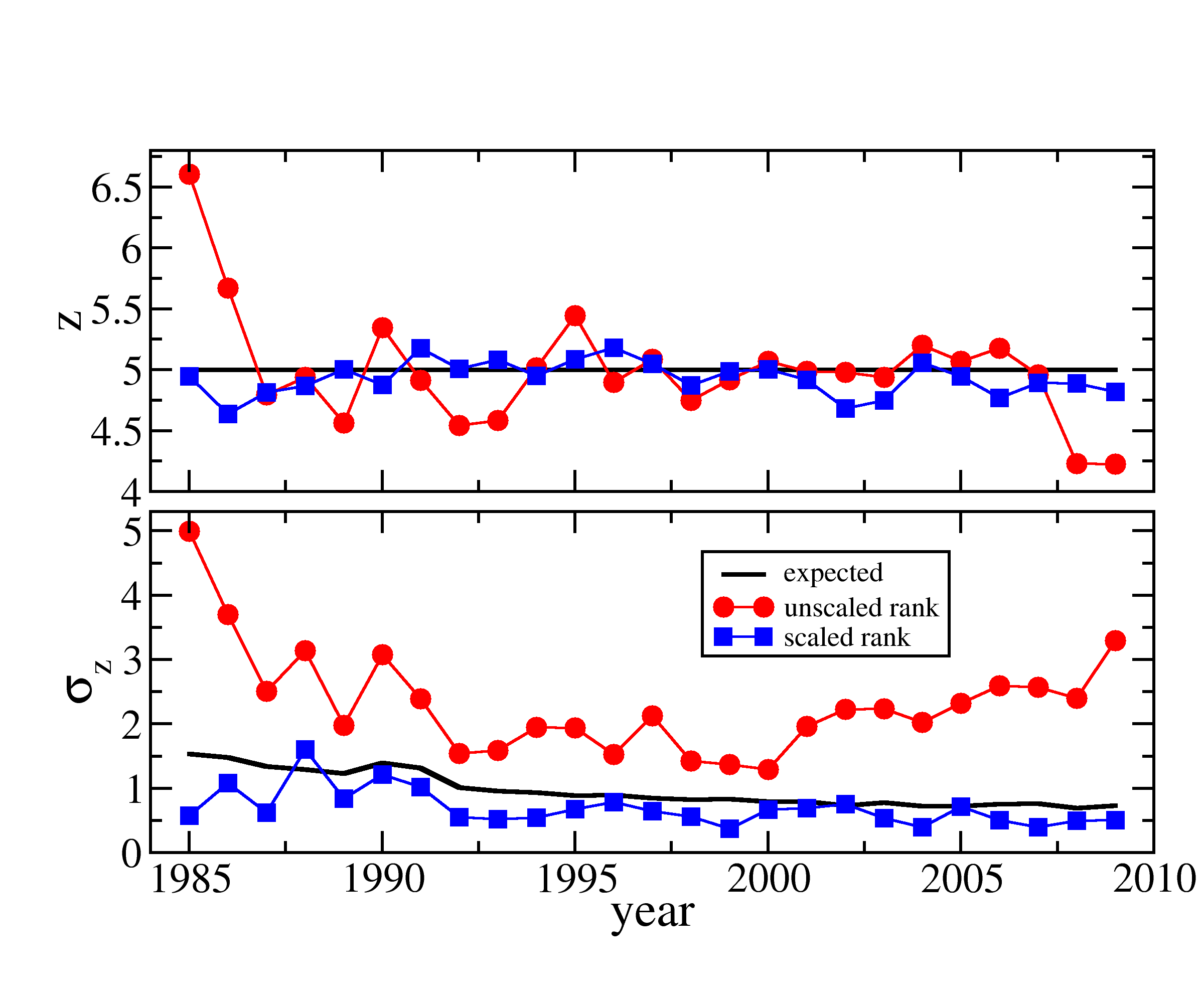}
\end{center}
\caption{
(Color online) Comparison between the theoretical values (black line) for the mean (top panel)
and standard deviation (bottom panel) of bins height in 
Fig.~\ref{z=5}  and
the values obtained when the ranking is performed based on $c$ (red circles)
or $c_f$ (blue squares).}
\label{table2}
\end{figure}
This quantity measures the success of a paper, in terms of citations received,
compared with other papers in the same category and year. A value
$c_f>1$ $(<1)$ indicates that the paper has been cited more (less) than the
average.
By definition the average value of $c_f$ is 1, for any category
or year, but this is not enough to have a useful unbiased indicator.
We need the full distributions to be the same, so that the probability of
having any value of $c_f$ is the same, no matter the category or the year.
In Fig.~\ref{distr_scaled} we show the distributions of
relative citation numbers for the same data of Fig.~\ref{distr_unscaled}.
The comparison between the two figures reveals that the normalization
procedure 
rescales very well all distributions on top of each other,
thus allowing a fair comparison among publications in different fields
and/or years.
Some variation shows up only for recent publication years ($\ge 2005$)
for which citation patterns are generally far from
stationary~\cite{stringer08}. But also in these cases the difference 
with respect to the asymptotic shape of the scaled distribution
is small.
The scaling procedure is able to remove the bias almost completely
also for papers published one year ago.
\begin{figure*}
\begin{center}
\includegraphics[width=0.7\textwidth]{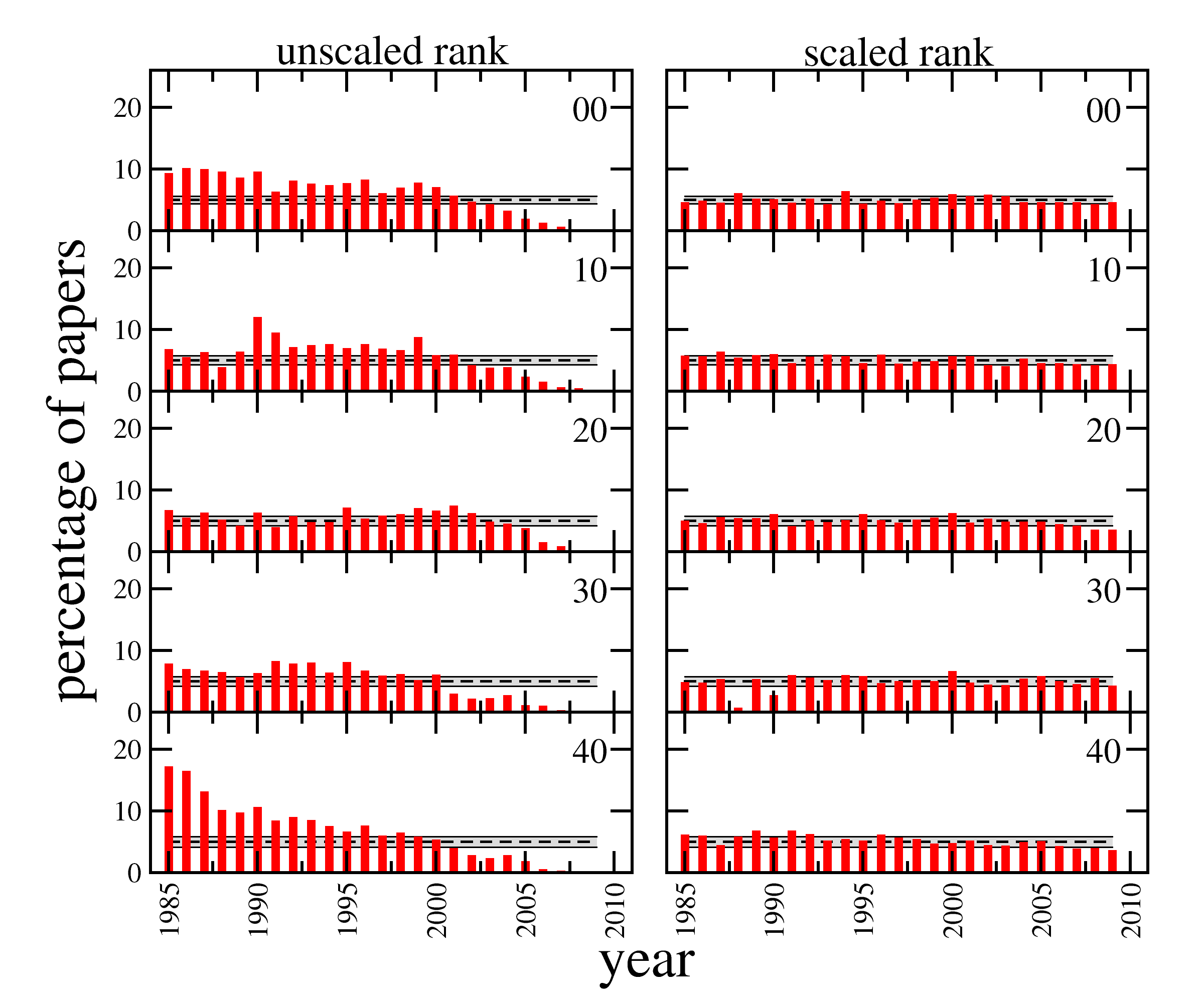}
\end{center}
\caption{(Color online) Histograms representing the percentage of papers belonging
to the top $5\%$ of the global rank performed according to $c$ (left
column) or $c_f$ (right column).
In each panel we consider only papers with a given PACS number
and plot the percentage of papers published in certain year belonging to
the top $5\%$ of the global rank. 
Black dashed lines represent the theoretically
expected values (in case of a fair ranking) of the mean, while
the gray areas cover the regions corresponding to the
mean $\pm$ one standard deviation.}
\label{z=5pacs}
\end{figure*}
\begin{figure*}
\begin{center}
\includegraphics[width=0.7\textwidth]{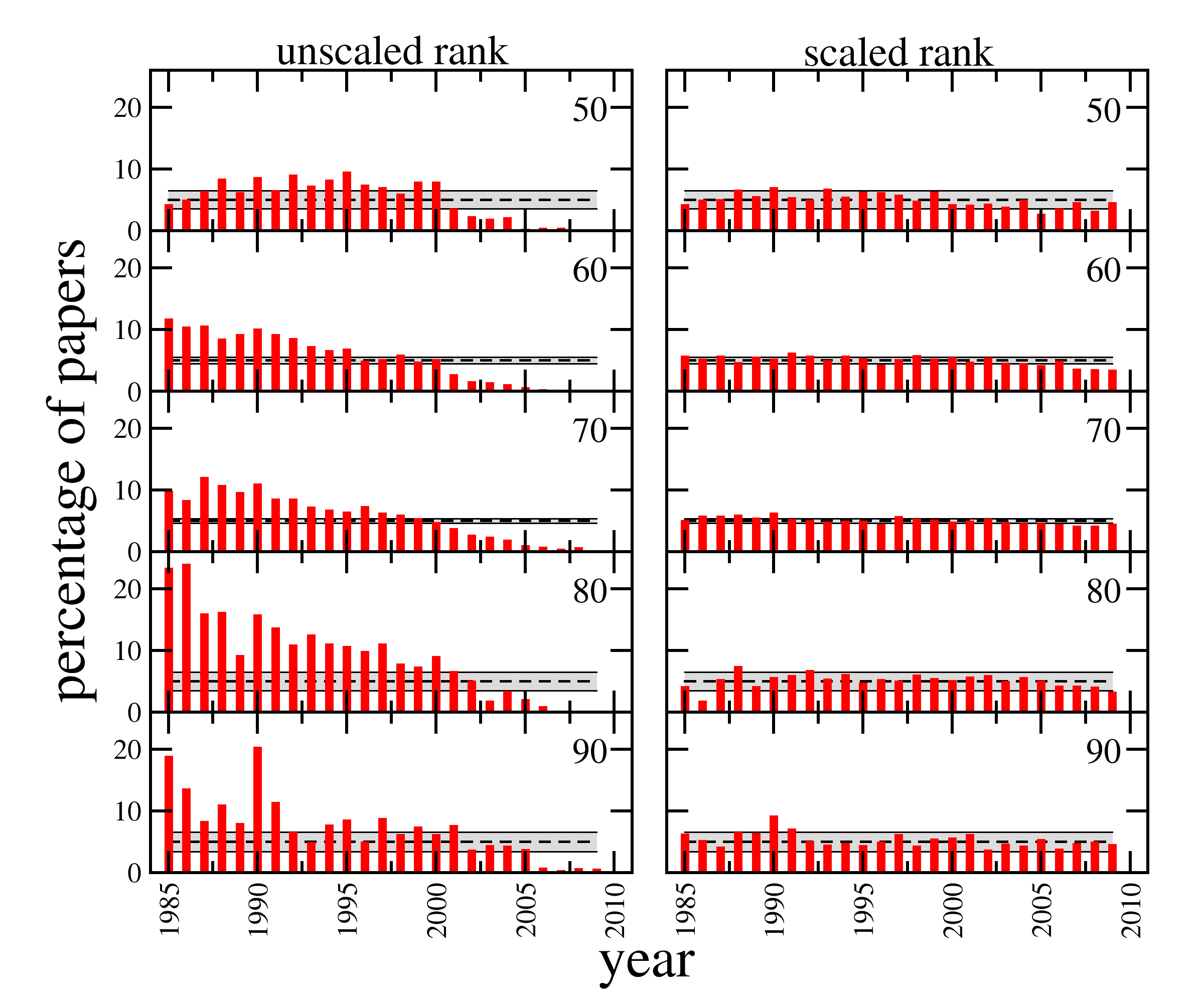}
\end{center}
\caption{Same as fig.~\ref{z=5pacs} but for the remaining set of PACS numbers.}
\label{z=5pacs_b}
\end{figure*}
\begin{figure}
\begin{center}
\includegraphics[width=0.45\textwidth]{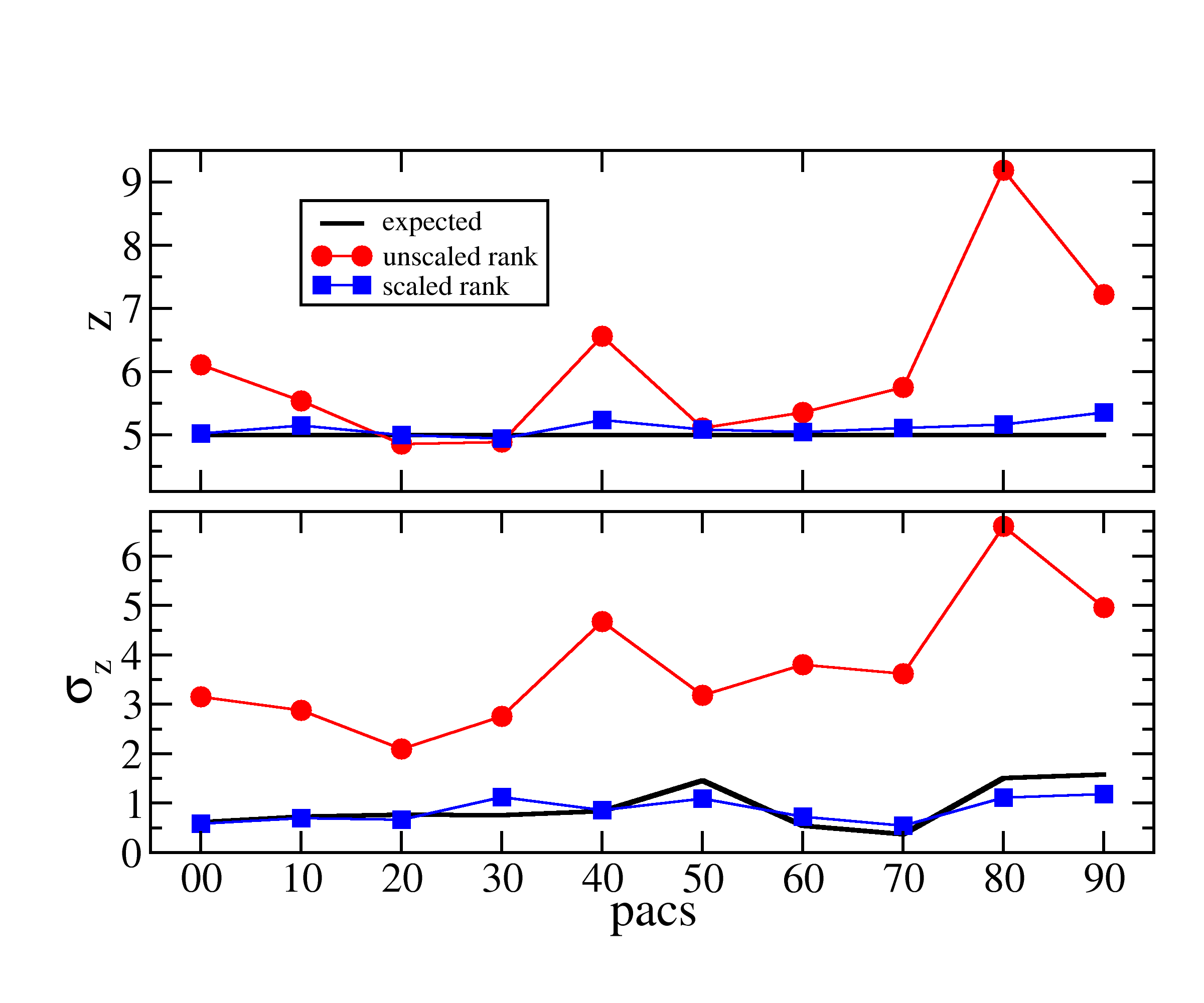}
\end{center}
\caption{
(Color online) Comparison between the theoretical values (black line) for the mean (top panel)
and standard deviation (bottom panel) of bins height in 
Fig.~\ref{z=5pacs}  and
the values obtained when the ranking is performed based on $c$ (red circles)
or $c_f$ (blue squares).}
\label{table3}
\end{figure}
A more quantitative assessment of the improvement induced by the use
of $c_f$ instead of the raw number of citations $c$ comes from
the ranking of articles.
We consider papers belonging to several sets
and rank them according either to the number of citations $c$ or to
the rescaled indicator $c_f$.
We compute then the percentage of publications of each category
that appear in the top $z\%$ of the global rank. If the ranking is fair
the percentage for each category should be around $z\%$ with small
fluctuations. 

Fig.~\ref{z=5} clearly shows that when articles are ranked
according to the unnormalized number of citations $c$ there are wide
variations among fields. Such variations are dramatically reduced
when the relative indicator $c_f$ is used.
More quantitatively, assuming that the distributions for the various
fields are the same, the expected value of the bin height in
Fig.~\ref{z=5} is $z\%$ with a standard deviation
\begin{equation}
\sigma_z = \sqrt{ \frac{z
  \left(100-z\right)}{N_c} \sum_{i=1}^{N_c} \frac{1}{N_i}},
\end{equation}
where $N_c$ is the number of categories and $N_i$ the number of articles
in the $i$-th category.
When the ranking is performed according to
$c_f=c/c_0$ we find (Fig.~\ref{table2}) a very good agreement with
the hypothesis that the ranking is unbiased, while strong evidence
that the ranking is biased is found when $c$ is used. 

We show in Figs.~\ref{z=5pacs},~\ref{z=5pacs_b} and~\ref{table3} the
results obtained for different publication years,
but fixed PACS categories.

The results presented in the figures demonstrate that the
ranking based on $c_f$ is fair, the fluctuations around the expected value
$z \%$ being accounted for by finite sample effects. On the contrary, when the
raw number $c$ of citations is used, some sets are largely overrepresented
and others underrepresented in the top $5 \%$ group.
Similar results (not shown) are obtained for other values of $z$.

\section{Renormalized citation counts for authors}
In this section we present a practical application
to the comparison of individual researchers, showing how the rescaled
indicator $c_f$ is correlated but far from 
being
equivalent to the raw number of citations $c$. 
We have considered all papers published in APS journals from $1985$ to
$2006$ and labeled  with PACS numbers,
and indicated, for the generic paper $i$, with
$c(i)$ and $c_f(i)$ the number of citations received 
and the relative indicator, respectively.
We have then identified, for each author $a$, the set $\{a\}$
of her/his publications~\footnote{This procedure is subject to many
potential errors in the identification of authors. However, this problem is not
big~\cite{radicchi09}.}, and computed
the total number of her/his citations $C^{(a)}=\sum_{i \in \{a\}} c(i)$
and the corresponding total value of the her/his relative indicator
$C^{(a)}_f = \sum_{i \in \{a\}} c_f(i)$.
\begin{figure}
\begin{center}
\includegraphics*[width=0.45\textwidth]{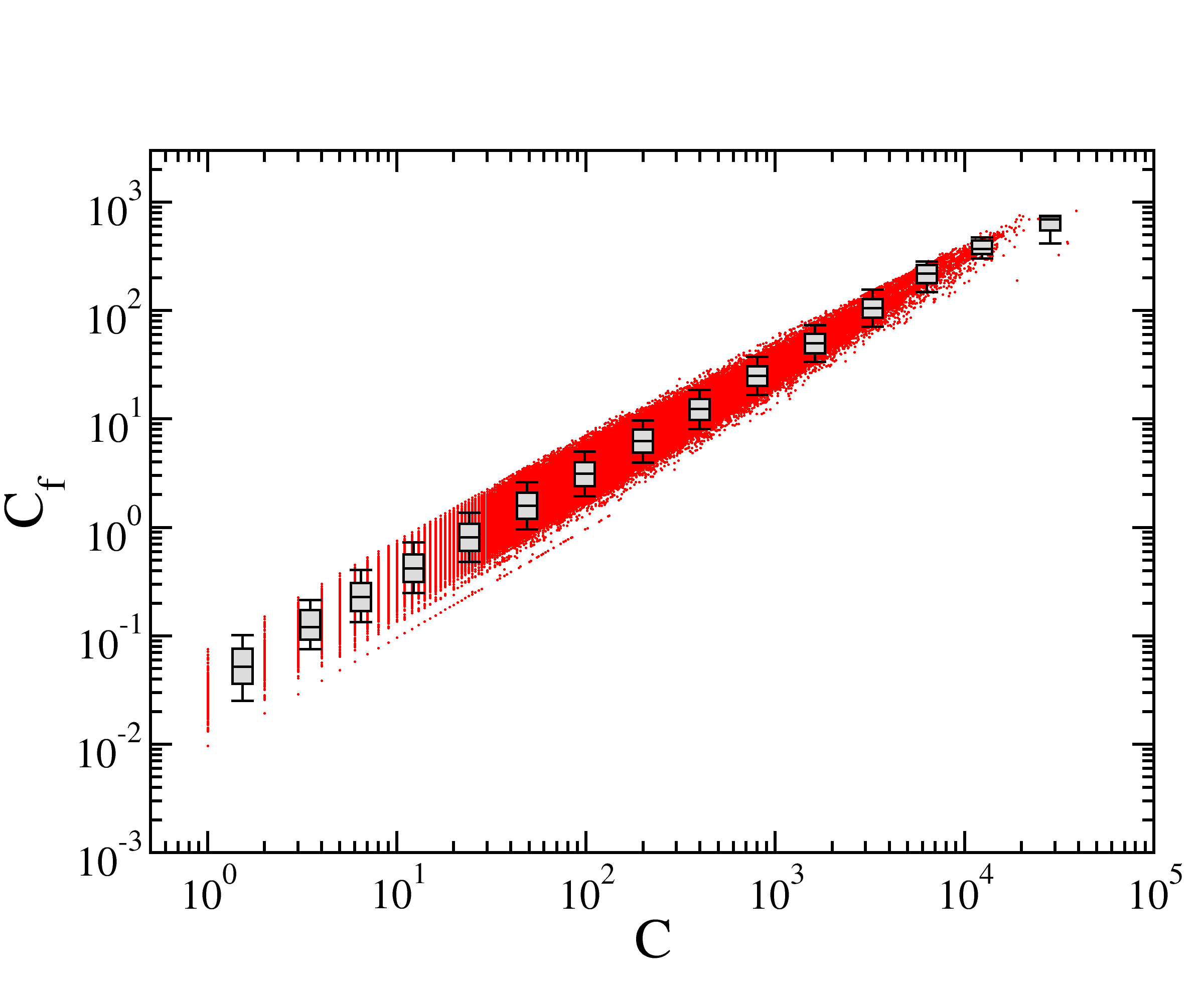}
\end{center}
\caption{(Color online) Scatter plot of the values of $C$ and $C_f$ for each author
who published in APS journals between $1985$ and $2006$.
For clarity, data have been logarithmically binned and for each bin we plot
the values of $c_f$ corresponding to the 
$90\%$ confidence intervals (error bars), $25\%$ confidence intervals 
(boxes) and median (horizontal bars inside boxes).
}
\label{example1}
\end{figure}
Fig.~\ref{example1} reports for each author the value of the total rescaled
indicator $C_f$ as a function of the total number of raw citations $C$.
The correlation is very good $r \approx 0.98$, but not perfect.
For a single value of one indicator there are often values of the other
spanning one order of magnitude, indicating again that authors with a very
different number of raw citations may be equivalent once differences
across subfields and time are considered.
\
An even more striking result is obtained when the relative indicator
is further rescaled by the number of authors for each paper:
$B_f^{(a)} = \sum_{i \in \{a\}} c_f(i)/N(i)$, where $N(i)$ is the number of authors
of the $i$-th paper.
This rescaling is aimed at taking into account multiple
authorship~\cite{price81, egghe00, hsu09} and it is based on the assumption
that all authors
contribute equally to each publication~\footnote{This is for sure largely
incorrect in many cases, but it is the most reasonable assumption
as long as precise statements about individual contributions are
not be published for each article.}.
\begin{figure}
\begin{center}
\includegraphics*[width=0.45\textwidth]{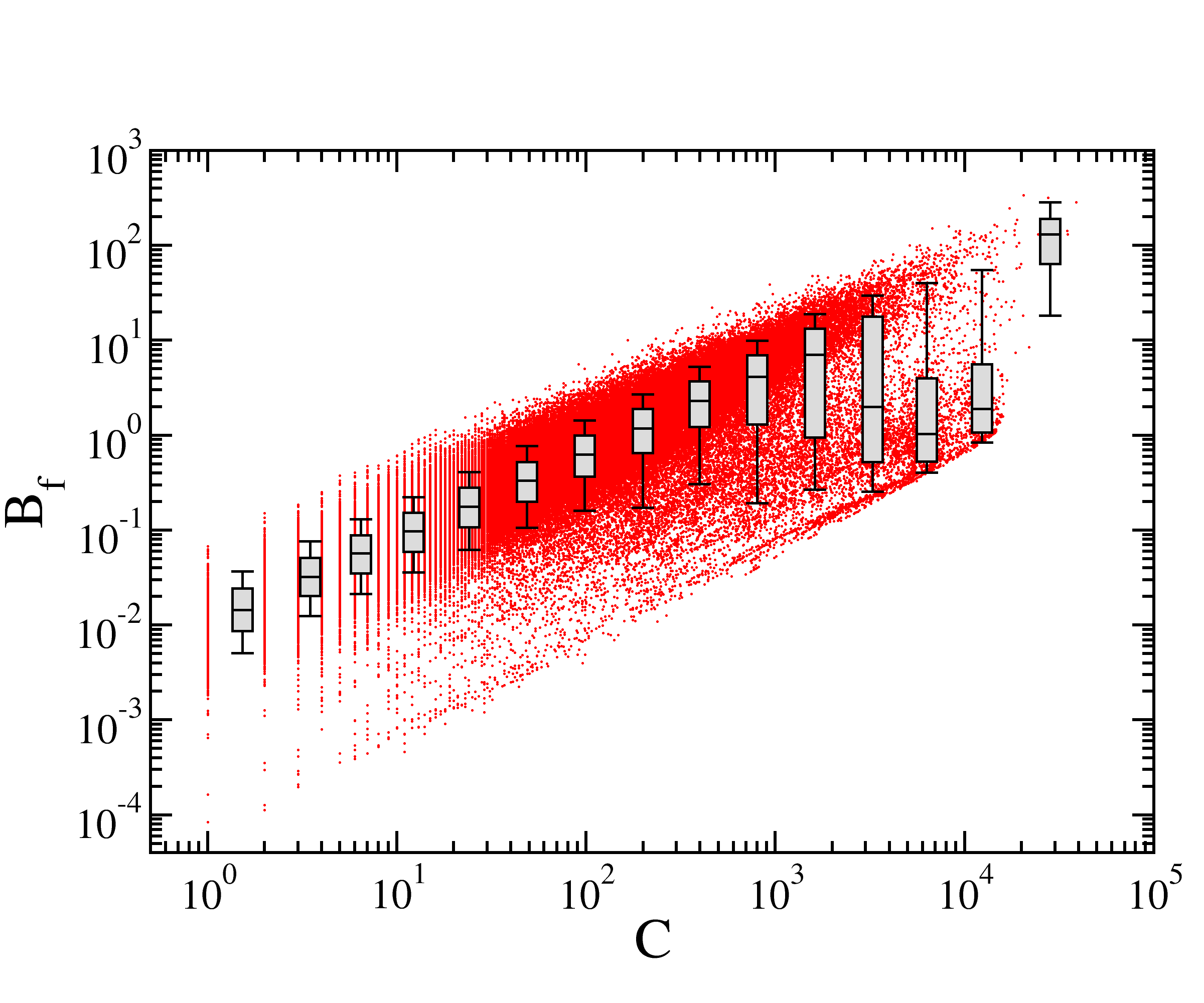}
\end{center}
\caption{(Color online) Scatter plot of the values of $C$ and $B_f$ for each author
who published in APS journals between $1985$ and $2006$.
For clarity, data have been logarithmically binned and for each bin we plot
the values of $B_f$ corresponding to the 
$90\%$ confidence intervals (error bars), $25\%$ confidence intervals 
(boxes) and median (horizontal bars inside boxes).
}
\label{example2}
\end{figure}
In this case, the range of variations of $B_f$ for each value of 
$C$ is much increased (the $r$-value is reduced to $r=0.49$),
reaching in many case two orders of magnitudes. Fig.~\ref{example2}
leads to the striking observation that a researcher with $10^4$
citations may have a number of rescaled citations equal to a colleague
whose publications have been cited only less than $100$ times!

\section{Conclusions}
In this paper we have investigated the possibility to compare in a
fair manner the citations of papers published in APS journals in
different years and/or in different fields of Physics.
We have shown that the raw number of citations is not a suitable indicator,
since there are remarkable differences depending on the field and on the year
of publication. A fair comparison is obtained instead if the
relative number of citations $c_f$ (i.e., the number of citations
divided by the average number
of cites for the same category and year) is considered. The normalization
rescales essentially all distributions on top of each other and this
is further confirmed by checking that ranking papers according to $c_f$
does not introduce any bias.
For completeness, we have performed (but not shown here) the
same type of analysis by using as renormalization factor 
the median value instead of the average. 
The median is less sensitive to possible extreme events such as the
presence of highly cited papers, but diving the raw number of cites 
by the median value leads to less fair comparisons and only for 
sufficiently old publications ($< 2000$).

In this paper we have considered fields as identified by the first digit
of the first number in the PACS code. This classifies all papers in Physics
in $10$ very broad categories. In principle one can pursue further
this line of investigation, considering more refined levels of categorization.
A natural next step would be the consideration of $100$ distinct categories,
each identified by the whole first field of the PACS code. However,
the number of papers published each year in each of these categories is
typically very small, and this gives rise to huge fluctuations that
do not allow to extract reliable conclusions.

We believe that these results are very important in view of the increasing
trend towards quantitative evaluation of research performance.
We strongly encourage researchers dealing with such issue to consider relative
citation numbers as the basis of all their evaluations.
All indicators for sets of publications (individual authors, groups, research
institutions) must be constructed based on the relative citation
numbers. This is important also for very large sets
(e.g., at the institution level) in order to weigh in a
balanced manner the contribution of all fields.
For this reason the values of the average number of citations $c_0$
for each category and each year will be available at the web page
{\tt filrad.homelinux.org/resources},
where they will be updated periodically.

APS journals are an important but clearly partial domain of the
whole range of dissemination outlets available for researchers
in Physics. The extension of the work presented here to include also
all other journals where research about Physics is published
is a much needed step toward a more reliable citation-based research
performance evaluation.
Finally, let us stress that the attribution of citations
of multi-authored papers to individual contributors is a crucial
and much overlooked issue. Different ways of dealing with this
problem lead to completely different results, as Figs.~\ref{example1}
and~\ref{example2} demonstrate. The current common habit of attributing
all citations to all authors, with no normalization, is unfair and it
encourages the misconduct of inflating author lists with 
individuals who did not actually contribute to the work.

\section{Acknowledgments}
We acknowledge the American Physical Society
for providing the data about Physical Review’s journals.
We thank L.A.N.~Amaral and M.~Cencini for discussions and feedback
on the manuscript.

\end{document}